\documentclass[a4paper,11pt]{article}
\usepackage{amsfonts}
\usepackage{amsmath}
\usepackage{amssymb}
\usepackage{graphicx}
\usepackage{array}
\usepackage{booktabs}
\usepackage{multirow}
\usepackage{makecell}
\usepackage{float}
\pdfoutput=1 

\usepackage{subcaption}
\usepackage{mymacros}
\usepackage{chngcntr}
\usepackage{verbatim}
\usepackage{amsthm}
\usepackage{graphics}
\usepackage{xcolor}
\usepackage{mathrsfs}
\usepackage{bbold}
\usepackage{cases}
\usepackage{epsfig}
\usepackage{epstopdf}
\usepackage{hyperref}
\usepackage{amsfonts}
\usepackage{hyperref}
\usepackage{jheppub} 

\usepackage[T1]{fontenc} 

\hypersetup{
	colorlinks=true,
	linkcolor=blue,
	filecolor=blue,
	urlcolor=blue,
	citecolor=blue,
}

\title{\boldmath Particle dynamics in rotating charged Bumblebee black holes: An analytical correspondence with Kerr–Newman}

\author{Xiaowei Li$^a$\footnote{E-mail: xiaoweiliedu@hotmail.com}, }
\author{Deyou Chen$^b$\footnote{E-mail: deyouchen@hotmail.com}, }
\author{Chuang Yang$^b$\footnote{E-mail: chuangyangyc@hotmail.com}}

\affiliation{$^a$ School of Mathematics, Physics and Statistics, Sichuan Minzu College, Kangding 626001, People’s Republic of China\\
$^b$ School of Science, Xihua University, Chengdu 610039, People’s Republic of China}

\abstract{In this paper, we establish an analytical correspondence between the equatorial dynamics of spinless and spinning test particles in a slowly rotating charged Einstein--Bumblebee black hole spacetime and those in an effective Kerr--Newman spacetime. For spinning particles, the correspondence is established within the Mathisson--Papapetrou--Dixon framework under appropriate parameter matching. We show that corresponding circular orbits have identical radii and coordinate angular velocities, while their	radial dynamics and local instability rates exhibit a Lorentz-violation-dependent scaling. Nevertheless, the correspondence does not imply complete dynamical equivalence, as orbital trajectories and 	radial perturbation growth over an azimuthal period	generally differ. These results delineate the scope and limitations of the effective Kerr--Newman correspondence by distinguishing the dynamical properties preserved under parameter matching from those modified by the Lorentz-violating radial deformation.}


\begin{document} 
	\maketitle
	\flushbottom
	
\section{Introduction}
\label{sec:introduction}

Spontaneous Lorentz symmetry breaking offers a possible mechanism for departures from local Lorentz invariance in gravitational theories. In Einstein--Bumblebee gravity, this mechanism is realized through a vector field that acquires a nonzero vacuum expectation value, thereby
selecting a preferred direction in spacetime. Through its nonminimal coupling to curvature, the vector field can modify the geometric structure of black hole solutions. Static, rotating, and charged black hole solutions, including extensions with a cosmological constant, have been investigated within this framework \cite{Casana2018,Ding2020,MalufNeves2021,DingChen2021,AraujoFilho2023,AraujoFilho2024}. Of particular interest are slowly rotating charged Bumblebee black holes \cite{LGWL}, whose metric exhibits a close structural relation to the Kerr--Newman (KN) geometry. For matched effective background parameters, their temporal--azimuthal metric sectors coincide, whereas their radial metric components differ by a factor involving the Lorentz-violating parameter. This geometric structure raises the question of whether the correspondence between the two metrics extends to test-particle dynamics and, if so, which dynamical properties are preserved and which are modified by the Lorentz-violating radial deformation.

The dynamics of test particles in KN spacetime has been extensively investigated, including analytical studies of spinless particle motion~\cite{Hackmann2013} and spinning particle dynamics~\cite{HH}. In Bumblebee black hole backgrounds, previous studies have examined massive particle deflection in rotating geometries~\cite{LiOvgun2020} and the orbital dynamics and stability of spinning particles in static charged configurations~\cite{Li2026}. In parallel, analytical methods have been developed to construct Bumblebee geometries from general-relativistic backgrounds~\cite{Poulis2022,Ovcharenko2026}. These studies provide important insights into particle motion and the geometric properties of Lorentz-violating black holes. However, they do not establish whether the equatorial dynamics of spinless and spinning particles in a rotating charged Bumblebee background admits an analytical correspondence with that in an effective KN spacetime, particularly when finite-spin effects are taken into account. Such an extension is nontrivial because the Mathisson--Papapetrou--Dixon (MPD) equations couple the particle spin directly to the curvature of the background spacetime. Under the Tulczyjew--Dixon (TD) spin supplementary condition, the four-momentum and four-velocity are generally not parallel, and a consistent mapping must account for both the effective background parameters and the particle spin. Moreover, an analytical correspondence does not necessarily imply complete dynamical equivalence. Even if corresponding circular orbits share the same radius and coordinate angular velocity, their radial evolution and orbital trajectories need not coincide because of the Lorentz-violating radial deformation. The radial Lyapunov exponent (LE) of an unstable circular orbit provides a quantitative measure of local orbital instability and thus allows us to examine whether the growth rates of radial perturbations are preserved under the correspondence~\cite{CMBWZ}. This also motivates comparing the LE with the black hole surface gravity.

In this work, we investigate the analytical correspondence between the equatorial dynamics of neutral spinless and spinning test particles in a slowly rotating charged Bumblebee black hole spacetime and those in an effective KN spacetime. Starting from the geometric relation between the two backgrounds, we first establish the correspondence for spinless geodesic motion and then extend the analysis to spinning particles within the MPD framework supplemented by the Tulczyjew--Dixon (TD) spin supplementary condition. By determining the appropriate matching of the effective background parameters and the signed particle spin, we examine which orbital properties are preserved under the correspondence and which are modified by the Lorentz-violating radial deformation. We further investigate the associated local radial instability through the LE and compare it with the black hole surface gravity, thereby clarifying the physical implications and limitations of the analytical correspondence.

The remainder of this paper is organized as follows. Section~\ref{sec2} introduces the rotating charged Bumblebee black hole background and its relation to an effective KN spacetime. Section~\ref{sec3} establishes the correspondence for spinless geodesic motion, while Section~\ref{sec4} extends the analysis to equatorial spinning-particle dynamics within the MPD--TD framework. Section~\ref{sec5} discusses the physical implications and limitations of the correspondence, including its application to the chaos-bound criterion. Finally, Section~\ref{sec6} presents our conclusions.

\section{Bumblebee--Kerr--Newman spacetime}
\label{sec2}

We consider a slowly rotating charged black hole in Einstein--Bumblebee gravity with a vanishing cosmological constant and a quadratic potential $V(X)=\lambda_B X^2/2$, where
$X=B^\mu B_\mu\pm b^2$ measures the deviation from the vacuum condition and $\lambda_B$ is
the potential coupling constant. The metric in Boyer--Lindquist coordinates is given by \cite{LGWL}

\begin{align}
ds^2={}&-\frac{\Delta_r}{\rho^2}\left(dt-a\sqrt{1+\ell}\sin^2\theta\,d\phi\right)^2
+(1+\ell)\frac{\rho^2}{\Delta_r}dr^2+\rho^2d\theta^2 \nonumber\\
&+\frac{\sin^2\theta}{\rho^2}\left(a\sqrt{1+\ell}\,dt-\left(r^2+a^2(1+\ell)\right)d\phi\right)^2,
\label{eq:bumblebee_metric}
\end{align}

\noindent where

\begin{align}
\Delta_r &=r^2-2Mr+(1+\ell)a^2 +\frac{2(1+\ell)Q_0^2}{2+\ell},\nonumber\\
\rho^2 &=r^2+(1+\ell)a^2\cos^2\theta.
\label{eq:bumblebee_metric_functions}
\end{align}

\noindent Here, $M$, $a$, and $Q_0$ denote the mass, rotation and charge parameters, respectively, while $\ell$ characterizes Lorentz-symmetry breaking. The corresponding electromagnetic four-potential reads

\begin{equation}
A_\mu= \left(-\frac{Q_0r}{\rho^2}, 0,0,\frac{Q_0a\sqrt{1+\ell}\,r\sin^2\theta}{\rho^2}\right).
\label{eq:bumblebee_potential}
\end{equation}

\noindent The metric reduces to the standard KN geometry as $\ell\to0$. For nonzero $\ell$, the rotating BKN background is an approximate solution whose field-equation residuals are of order $\mathcal O(a^2\ell)$ \cite{LGWL}. We therefore impose $a^2|\ell|/M^2\ll1$ and $\ell>-1$.
All subsequent geometric and particle-dynamical relations are understood within the validity
of this background approximation.

To relate the Bumblebee--Kerr--Newman (BKN) geometry to the standard KN spacetime, we introduce  the effective rotation and charge parameters

\begin{equation}
\bar a=a\sqrt{1+\ell}, \qquad \bar Q^2= \frac{2(1+\ell)}{2+\ell}Q_0^2.
\label{eq:effective_parameters}
\end{equation}

\noindent The metric functions then take the KN form,

\begin{equation}
\Delta_r=r^2-2Mr+\bar a^2+\bar Q^2, \qquad \rho^2=r^2+\bar a^2\cos^2\theta.
\label{eq:effective_metric_functions}
\end{equation}

\noindent For matched effective parameters, the two metrics satisfy

\begin{equation}
g_{\mu\nu}^{\rm BB} = g_{\mu\nu}^{\rm KN}(M,\bar a,\bar Q) +\ell\frac{\rho^2}{\Delta_r}
\delta_\mu^r\delta_\nu^r,
\label{eq:metric_mapping}
\end{equation}

\noindent where $g_{\mu\nu}^{\rm KN}$ denotes the standard KN metric in the same
Boyer--Lindquist coordinates. Thus, all covariant metric components coincide except for the radial component,

\begin{equation}
g_{rr}^{\rm BB}=(1+\ell)g_{rr}^{\rm KN}.
\label{eq:radial_metric_relation}
\end{equation}

\noindent The effective charge $\bar Q$ enters the metric functions, whereas the electromagnetic potential in Eq.~\eqref{eq:bumblebee_potential} depends on the original charge parameter $Q_0$. The metric correspondence therefore does not, in general, extend to the
electromagnetic field of a KN black hole with charge $\bar Q$. Moreover, Eq.~\eqref{eq:metric_mapping} is an exact algebraic identity within the adopted background, rather than an isometry between the two spacetimes. The radial deformation generally modifies the connection and curvature tensors, which are relevant to the spin--curvature coupling in the MPD equations.

The inner and outer horizons are determined by $\Delta_r(r_\pm)=0$, yielding

\begin{equation}
r_\pm=M\pm\sqrt{M^2-\bar a^2-\bar Q^2}.
\label{eq:BKN_horizons}
\end{equation}

\noindent  Here, $r_+$ and $r_-$ denote the event and inner horizons, respectively. We restrict attention to nonextremal configurations satisfying $\bar a^2+\bar Q^2<M^2$. For matched effective parameters, the BKN and effective KN black holes share the same horizon radii and angular velocity, $\Omega_H=\bar a/(r_+^2+\bar a^2)$, whereas their surface gravities satisfy

\begin{equation}
\kappa_{\rm BB} =\frac{\kappa_{\rm KN}}{\sqrt{1+\ell}}	=\frac{r_+-r_-}	{2\sqrt{1+\ell}(r_+^2+\bar a^2)}.
\label{eq:kappa_mapping}
\end{equation}

\noindent Here, $\kappa_{\rm BB}$ and $\kappa_{\rm KN}$ denote the surface gravities of the Bumblebee and effective KN black holes, respectively. Although the matched backgrounds have identical horizon radii and angular velocities, their surface gravities differ due to the radial metric deformation. Both vanish in the extremal limit.

\section{Analytical correspondence in the spinless limit}
\label{sec3}

Before considering spinning particles, we establish the analytical correspondence for neutral spinless geodesics in the BKN and effective KN backgrounds. This provides a reference for identifying the additional effects of spin--curvature coupling. We consider equatorial motion,
$\theta=\pi/2$, and normalize the particle rest mass to unity. Stationarity and axial
symmetry imply the conserved energy and angular momentum $E_0=-p_t$ and $J_0=p_\phi$, respectively. The subscript $0$ denotes spinless particle quantities, while $r_0$ denotes a circular-orbit radius. Throughout this section, the backgrounds are compared at fixed effective parameters $(M,\bar a,\bar Q)$ and identical particle constants $(E_0,J_0)$.

The four-momentum of a unit-mass spinless particle satisfies the mass-shell condition
$g^{\mu\nu}p_\mu p_\nu=-1$. On the equatorial plane, where $p_\theta=0$, this condition reduces to

\begin{equation}
g^{tt}E_0^2-2g^{t\phi}E_0J_0 +g^{\phi\phi}J_0^2+g^{rr}p_r^2=-1.
\label{eq:spinless_mass_shell}
\end{equation}

\noindent Using the metric correspondence in Eq.~\eqref{eq:metric_mapping}, the temporal--azimuthal contribution takes the form

\begin{equation}
g^{tt}E_0^2-2g^{t\phi}E_0J_0 +g^{\phi\phi}J_0^2 = -\frac{P_0^2(r)}{r^2\Delta_r}
+\frac{(J_0-\bar a E_0)^2}{r^2},
\label{eq:spinless_tphi_relation}
\end{equation}

\noindent with $P_0(r)=E_0(r^2+\bar a^2)-\bar a J_0$. Since the radial inverse metric component is $g_{\rm BB}^{rr}=\Delta_r/((1+\ell)r^2)$ and $p_r=g_{rr}\,dr/d\tau$, substitution
into Eq.~\eqref{eq:spinless_mass_shell} gives

\begin{equation}
\left(\frac{dr}{d\tau}\right)_{\rm BB}^2=\frac{R_0(r)}{(1+\ell)r^4},
\label{eq:spinless_radial_BB}
\end{equation}

\noindent where

\begin{equation}
R_0(r)=P_0^2(r)-\Delta_r\left(r^2+(J_0-\bar a E_0)^2\right).
\label{eq:spinless_R}
\end{equation}

\noindent The charge parameter enters through $\Delta_r$, while the factor $(1+\ell)^{-1}$
arises from the radial metric deformation. For the effective KN background, the corresponding radial equation is \cite{Carter1968}

\begin{equation}
\left(\frac{dr}{d\tau}\right)_{\rm KN}^2 =\frac{R_0(r)}{r^4}.
\label{eq:spinless_radial_KN}
\end{equation}

\noindent These two systems therefore share the same radial function, with

\begin{equation}
\left(\frac{dr}{d\tau}\right)_{\rm BB} = \frac{1}{\sqrt{1+\ell}} \left(\frac{dr}{d\tau}\right)_{\rm KN},
\label{eq:spinless_proper_radial_mapping}
\end{equation}

\noindent for corresponding radial branches with the same orientation.

The temporal and azimuthal components of the four-velocity follow from $u^\mu=g^{\mu\nu}p_\nu$, which can be written as follows

\begin{align}
u^t &= 
\frac{1}{r^2} \left(\frac{(r^2+\bar a^2)P_0(r)}{\Delta_r}+\bar a(J_0-\bar a E_0)\right),
\label{eq:spinless_ut}\\
u^\phi &=
\frac{1}{r^2}\left(\frac{\bar a P_0(r)}{\Delta_r}+(J_0-\bar a E_0)\right).
\label{eq:spinless_uphi}
\end{align}

\noindent Because the temporal--azimuthal metric sectors coincide, these components are
identical in the matched backgrounds: $u_{\rm BB}^t=u_{\rm KN}^t$ and $u_{\rm BB}^\phi=u_{\rm KN}^\phi$. Consequently, the coordinate angular velocities satisfy

\begin{equation}
\Omega_{0,\rm BB}=\Omega_{0,\rm KN},
\label{eq:spinless_angular_velocity_mapping}
\end{equation}

\noindent where $\Omega_0=\frac{d\phi}{dt}=\frac{\bar a P_0(r)	+\Delta_r(J_0-\bar a E_0)}{	(r^2+\bar a^2)P_0(r)+\bar a\Delta_r(J_0-\bar a E_0)}$ is defined wherever $u^t\neq0$. Combining Eq.~\eqref{eq:spinless_proper_radial_mapping} with the equality of $u^t$ yields

\begin{equation}
\left(\frac{dr}{dt}\right)_{\rm BB}	=\frac{1}{\sqrt{1+\ell}} \left(\frac{dr}{dt}\right)_{\rm KN}.
\label{eq:spinless_coordinate_velocity_mapping}
\end{equation}

\noindent Hence, along corresponding noncircular segments with $dr/dt\neq0$,

\begin{equation}
\left(\frac{d\phi}{dr}\right)_{\rm BB}
=
\sqrt{1+\ell} \left(\frac{d\phi}{dr}\right)_{\rm KN}.
\label{eq:spinless_azimuthal_radial_mapping}
\end{equation}

\noindent For segments with the same radial endpoints and orientation, integration gives

\begin{equation}
\Delta\phi_{\rm BB} = \sqrt{1+\ell}\,\Delta\phi_{\rm KN},
\label{eq:spinless_accumulated_azimuth}
\end{equation}

\noindent where $\Delta\phi$ denotes the accumulated coordinate azimuthal angle. Thus, equal
coordinate angular velocities do not generally imply identical trajectories in the $(r,\phi)$ plane.

For a circular orbit at $r=r_0>r_+$, the radial function satisfies

\begin{equation}
R_0(r_0)=0, \qquad R_0'(r_0)=0,
\label{eq:spinless_circular}
\end{equation}

\noindent where derivatives are taken at fixed $E_0$ and $J_0$. A regular circular orbit is radially unstable when $R_0''(r_0)>0$. Since the radial function is common to both backgrounds, corresponding physically admissible circular orbits have identical values of $(r_0,E_0,J_0)$. To determine the local instability rate, we introduce a radial perturbation $r(t)=r_0+\epsilon(t)$ with $|\epsilon|\ll r_0$. In coordinate time, the BKN radial equation reads

\begin{equation}
\left(\frac{dr}{dt}\right)_{\rm BB}^2 = \frac{R_0(r)} {(1+\ell)r^4(u^t)^2}.
\label{eq:spinless_coordinate_radial}
\end{equation}

\noindent Expanding around the circular orbit,
using Eq.~\eqref{eq:spinless_circular},
gives

\begin{equation}
\left(\frac{d\epsilon}{dt}\right)^2= \frac{R_0''(r_0)}{2(1+\ell)r_0^4(u_0^t)^2}\epsilon^2
+\mathcal O(\epsilon^3),
\label{eq:spinless_perturbation_expansion}
\end{equation}

\noindent where $u_0^t=u^t(r_0)$ is assumed
finite and nonzero. The corresponding linearized radial
equation is

\begin{equation}
\frac{d^2\epsilon}{dt^2}
-\lambda_{0,\rm BB}^2\epsilon=0,
\label{eq:spinless_linear_perturbation}
\end{equation}

\noindent The squared coordinate-time LE of an unstable circular orbit in the BKN background is therefore

\begin{equation}
\lambda_{0,\rm BB}^2 = \frac{R_0''(r_0)} {2(1+\ell)r_0^4(u_0^t)^2}.
\label{eq:spinless_LE_BB}
\end{equation}

\noindent For the corresponding orbit in the effective KN background, the radial function $R_0(r)$, circular-orbit radius $r_0$, and temporal four-velocity $u_0^t$ are unchanged, whereas the radial rescaling factor is absent. Hence,

\begin{equation}
\lambda_{0,\rm KN}^2 = \frac{R_0''(r_0)}{2r_0^4(u_0^t)^2},
\qquad
\lambda_{0,\rm BB}^2 =	\frac{\lambda_{0,\rm KN}^2}{1+\ell}.
\label{eq:spinless_LE_mapping}
\end{equation}

\noindent The spinless correspondence thus preserves the circular-orbit conditions and coordinate angular velocities, while rescaling the radial coordinate velocities and LEs. These relations hold for matched effective background parameters and particle constants, rather than for a variation of $\ell$ at fixed original parameters $(a,Q_0)$. For spinning test particles, the MPD equations introduce spin--curvature coupling and a nontrivial momentum--velocity relation, so the spinless correspondence does not automatically extend to finite spin. This extension is examined in the next section.

\section{Spinning particle dynamics and effective Kerr–Newman correspondence}
\label{sec4}

\subsection{Equations of motion and conserved quantities}
\label{sec4.1}

We consider the motion of a neutral spinning test particle in the charged rotating Bumblebee spacetime. The dynamics of spinning test particles in rotating black hole spacetimes has been investigated within the Papapetrou framework, including studies of orbital stability and chaotic motion~\cite{Hartl2003}. Within the pole--dipole approximation, the dynamics of a spinning test particle is governed by the MPD equations \cite{HH,Mathisson1937,Papapetrou1951,Dixon1970}.

\begin{align}
\frac{D\tilde{p}^{\mu}}{D\tau}
&=-\frac{1}{2}R^{\mu}{}_{\nu\alpha\beta}
u^{\nu}\tilde{S}^{\alpha\beta},
\label{mpd1}\\
\frac{D\tilde{S}^{\mu\nu}}{D\tau}
&=\tilde{p}^{\mu}u^{\nu}
-u^{\mu}\tilde{p}^{\nu},
\label{mpd2}
\end{align}

\noindent where $D/D\tau$ denotes the covariant derivative along the particle worldline, parametrized by $\tau$. The quantities $\tilde{p}^{\mu}$ and $u^{\mu}=dx^{\mu}/d\tau$ denote the four-momentum and four-velocity of the particle, respectively. Here, $R^{\mu}{}_{\nu\alpha\beta}$ is the Riemann curvature tensor, while $\tilde{S}^{\mu\nu}$ represents the antisymmetric spin tensor. We take $\tau$ to be the proper time along the	representative worldline, so that $u^\mu u_\mu=-1$. Under the TD condition, the four-momentum and four-velocity are generally not parallel because of spin--curvature coupling, and their relation must be determined consistently within the pole--dipole approximation~\cite{Han2010,ZhangLiu2019}. To close the MPD system, a spin supplementary condition is required to specify the representative worldline of the spinning particle. We adopt the TD condition \cite{WT1,WT2},

\begin{equation}
\tilde{S}^{\mu\nu}\tilde{p}_{\nu}=0.
\label{TD}
\end{equation}

\noindent Under this condition, the dynamical mass $m=\sqrt{-\tilde{p}^{\mu}\tilde{p}_{\mu}}$
and spin magnitude $\tilde{S}=\sqrt{\tilde{S}_{\mu\nu}\tilde{S}^{\mu\nu}/2}$ are conserved within the pole--dipole approximation. For convenience, we introduce the mass-normalized
quantities

\begin{equation}
p^{\mu}=\frac{\tilde{p}^{\mu}}{m},
\qquad
S^{\mu\nu}=\frac{\tilde{S}^{\mu\nu}}{m},
\qquad
S=\pm\frac{\tilde{S}}{m},
\label{eq:normalized_variables1}
\end{equation}

\noindent which satisfy

\begin{align}
S^2&=\frac{1}{2}S_{\mu\nu}S^{\mu\nu},
\label{eq8}\\
-1&=p^{\mu}p_{\mu}.
\label{eq9}
\end{align}

\noindent The sign of $S$ specifies the spin orientation relative to the symmetry axis: $S>0$ corresponds to alignment with the positive $z$ direction, whereas $S<0$ denotes anti-alignment.

Following the standard treatment of spinning particle dynamics \cite{ZHA,HH1,HH3}, we restrict the motion to the equatorial plane, $\theta=\pi/2$, with the particle spin aligned or anti-aligned with the symmetry axis. We impose $p^\theta=u^\theta=0$ and $S^{\theta\mu}=0$, so that the potentially nonvanishing components of the spin tensor are $S^{tr}$, $S^{t\phi}$, and $S^{r\phi}$. In terms of the normalized variables, the TD condition takes the form $S^{\mu\nu}p_\nu=0$, yielding

\begin{align}
	p_rS^{tr}+p_\phi S^{t\phi}&=0,
	\label{eq:TD_component1}\\
	p_tS^{tr}-p_\phi S^{r\phi}&=0,
	\label{eq:TD_component2}\\
	p_tS^{t\phi}+p_rS^{r\phi}&=0.
	\label{eq:TD_component3}
\end{align}

\noindent These relations, together with the spin-magnitude and mass-shell constraints and the equatorial metric identities, $g_{t\phi}^2-g_{tt}g_{\phi\phi}=\Delta_r$ and $g_{rr}=\frac{(1+\ell)r^2}{\Delta_r}$, determine the spin-tensor components as

\begin{align}
S^{tr}=-2\sigma p_\phi, \quad 	S^{t\phi}=2\sigma p_r, \quad  S^{r\phi}&=-2\sigma p_t,
\label{eq2.10}
\end{align}

\noindent where $\sigma=\frac{S}{2r\sqrt{1+\ell}}$ is defined in terms of the signed normalized spin $S$. These expressions remain well defined at $p_r=0$ or $p_t=0$ and reduce to their KN counterparts in the limit $\ell=0$.

The stationarity and axisymmetry of the spacetime give rise to two conserved quantities associated with the Killing vectors $\xi_{(t)}^\mu=(\partial/\partial t)^\mu$ and $\xi_{(\phi)}^\mu=(\partial/\partial\phi)^\mu$. The conserved energy $E$ and total angular
momentum $J$, both normalized by the particle mass, are given by

\begin{align}
E&=-p_t+\sigma p_tg'_{t\phi}-\sigma p_\phi g'_{tt},
\label{eq2.13}\\
J&=p_\phi+\sigma p_\phi g'_{t\phi}-\sigma p_tg'_{\phi\phi},
\label{eq2.14}
\end{align}

\noindent where a prime denotes differentiation with respect to $r$. The conserved angular momentum $J$ contains both orbital and spin-dependent contributions. For $\bar a>0$, we distinguish the positive- ($J>0$) and negative- ($J<0$) angular-momentum sectors. Owing to the spin-dependent contributions and frame-dragging effects, the sign of $J$ does not necessarily determine the direction of orbital rotation in the chosen coordinates, which is characterized by the sign of $\Omega=d\phi/dt$. Using the effective parameters $\bar a$ and $\bar Q$ defined in Eq.~\eqref{eq:effective_parameters}, we introduce $h(r)=2M/r-\bar Q^2/r^2$, with $h'=-2M/r^2+2\bar Q^2/r^3$, where a prime denotes differentiation with respect to $r$. Substituting the equatorial metric components into Eqs.~\eqref{eq2.13} and \eqref{eq2.14} and solving for the covariant momentum components, we obtain

\begin{align}
p_t &=-\frac{(1-\bar a\sigma h')E+\sigma h'J}{\mathcal D_S},
\label{eq2.17}\\
p_\phi &=\frac{	(1+\bar a\sigma h')J-\sigma(2r+\bar a^2h')E}{\mathcal D_S},
\label{eq2.18}
\end{align}

\noindent where $\mathcal D_S=1+2r\sigma^2h'$. These expressions hold wherever $\mathcal D_S\neq0$. For equatorial motion, the mass-shell condition $p^\mu p_\mu=-1$ yields

\begin{align}
p_r^2 &=\frac{(1+\ell)r^2}{\Delta_r^2}(\mathcal P-\Delta_r),
\label{eq:BB_pr}\\
(p^r)^2&=\frac{\mathcal P-\Delta_r}{(1+\ell)r^2},
\label{eq:BB_pr_contravariant}
\end{align}

\noindent where 

\begin{equation}
 \mathcal P=g_{\phi\phi}p_t^2 -2g_{t\phi}p_tp_\phi +g_{tt}p_\phi^2.
\label{eq:p}
\end{equation}

To examine the extension of the effective KN correspondence to spinning-particle dynamics, we first consider the momentum relations between the two systems. As established in Eq.~\eqref{eq:metric_mapping}, the metrics differ only in their radial components, with $g_{rr}^{\rm BB}=(1+\ell)g_{rr}^{\rm KN}$, for matched effective parameters $(M,\bar a,\bar Q)$. The spin coefficients are $\sigma_{\rm BB}=S_{\rm BB}/(2r\sqrt{1+\ell})$ and $\sigma_{\rm KN}=S_{\rm KN}/(2r)$. Their equality requires the signed-spin rescaling

\begin{equation}
S_{\rm KN} =\frac{S_{\rm BB}}{\sqrt{1+\ell}}.
\label{eq:spin_mapping}
\end{equation}

\noindent This mapping matches the spin-dependent coefficients of the two systems and does not,
in general, identify particles with the same spin magnitude. For matched effective parameters, conserved quantities $(E,J)$, and signed spins related by Eq.~\eqref{eq:spin_mapping}, Eqs.~\eqref{eq2.17} and \eqref{eq2.18} yield

\begin{equation}
p_t^{\rm BB}=p_t^{\rm KN}, \qquad p_\phi^{\rm BB}=p_\phi^{\rm KN}.
\label{eq:covariant_momentum_mapping}
\end{equation}

\noindent Since the temporal--azimuthal blocks of the inverse metrics coincide, we further obtain

\begin{equation}
p_{\rm BB}^t=p_{\rm KN}^t, \qquad p_{\rm BB}^\phi=p_{\rm KN}^\phi.
\label{eq:contravariant_momentum_mapping}
\end{equation}

\noindent The radial mass-shell equations, together with the choice of corresponding radial
branches, give

\begin{equation}
p_{\rm BB}^r =\frac{p_{\rm KN}^r}{\sqrt{1+\ell}}, \qquad
p_r^{\rm BB} =\sqrt{1+\ell}\,p_r^{\rm KN}.
\label{eq:radial_momentum_mapping_full}
\end{equation}

\noindent These relations establish an algebraic correspondence between the four-momentum
components under the specified parameter matching. In particular, the corresponding radial mass-shell functions have identical zeros. However, the momentum correspondence does not imply equivalence of the spinning-particle trajectories. Under the TD condition, the four-momentum and four-velocity are generally not parallel owing to spin--curvature coupling.
The coordinate velocities $dr/dt$ and $d\phi/dt$ must therefore be examined separately using the MPD equations before establishing the correspondence between orbital LEs.

To determine the coordinate velocities, we combine the MPD equations \eqref{mpd1} and \eqref{mpd2} with the equatorial spin configuration. The spin tensor admits the dual representation

\begin{equation}
S^{ab}=-S\varepsilon^{abc}p_c, \qquad a,b,c\in\{t,r,\phi\},
\label{eq:BB_spin_dual}
\end{equation}

\noindent where $\varepsilon^{abc}$ is the Levi-Civita tensor of the induced equatorial metric, with $\varepsilon^{tr\phi}=1/(r\sqrt{1+\ell})$. Using the conservation of the spin magnitude and the covariant constancy of the induced Levi-Civita tensor, we obtain

\begin{equation}
\frac{DS^{ab}}{Dt} =-S\varepsilon^{abc}\frac{Dp_c}{Dt}.
\label{eq:BB_spin_derivative}
\end{equation}

\noindent Taking coordinate time $t$ as the evolution parameter, the MPD momentum equation becomes $Dp_\mu/Dt=-R_{\mu\nu\alpha\beta}(dx^\nu/dt)S^{\alpha\beta}/2$. Combining this relation with the spin-evolution equation yields

\begin{align}
p^t\dot r-p^r &=\sigma\left(\mathcal R_{\phi t}+\dot r\,\mathcal R_{\phi r}\right), \label{eq2.21}\\
p^t\dot\phi-p^\phi &=-\sigma\left( \mathcal R_{rt} +\dot\phi\,\mathcal R_{r\phi} \right),
\label{eq2.22}
\end{align}

\noindent where $\dot r=dr/dt$, $\dot\phi=d\phi/dt$, and $\mathcal R_{\mu\nu} =R_{\mu\nu\alpha\beta}S^{\alpha\beta}=4\sigma\left(-p_\phi R_{\mu\nu tr}+p_r R_{\mu\nu t\phi}-p_t R_{\mu\nu r\phi}\right)$. Solving Eqs.~\eqref{eq2.21} and \eqref{eq2.22} for the coordinate velocities and evaluating the corresponding curvature components of the charged rotating Bumblebee geometry, we obtain

\begin{align}
\dot r&=\frac{b_1}{a_1}, \label{eq2.23}\\
\dot\phi&=\frac{b_2}{a_2}. \label{eq2.24}
\end{align}

\noindent Here, the subscripts $1$ and $2$ label the auxiliary coefficients associated with the radial and azimuthal coordinate velocities, respectively. The auxiliary functions entering Eqs.~\eqref{eq2.23} and \eqref{eq2.24} are given by

\begin{align}
a_1={}&a_2 =p^t+\frac{S^2}{2(1+\ell)r^3} \left(\mathcal A_1p^t +\bar a(r^2+\bar a^2)\mathcal B_1p^\phi \right),
\label{eq:a1}\\
b_1 ={}&p^r\left(1-\frac{MS^2}{(1+\ell)r^3}+\frac{\bar Q^2S^2}{(1+\ell)r^4}\right),
\label{eq:b1}\\
b_2 ={}&p^\phi+\frac{S^2}{2(1+\ell)r^3} \left(-\bar a\mathcal B_1p^t +\mathcal C_1p^\phi \right),
\label{eq:b2}
\end{align}

\noindent where

\begin{align}
\mathcal A_1
={}&-2M+\frac{2\bar Q^2}{r}
-\frac{6M\bar a^2}{r^2}
+\frac{8\bar a^2\bar Q^2}{r^3},
\\
\mathcal B_1
={}&\frac{6M}{r^2}
-\frac{8\bar Q^2}{r^3},
\\
\mathcal C_1
={}&4M-\frac{6\bar Q^2}{r}
+\frac{6M\bar a^2}{r^2}
-\frac{8\bar a^2\bar Q^2}{r^3}.
\end{align}

\noindent These expressions reduce to $a_1=a_2=p^t$, $b_1=p^r$ and $b_2=p^\phi$ in the spinless limit $S\to0$. The radial coordinate velocity $\dot r=b_1/a_1$ will be used below
to determine circular orbits and their local radial instability.

\subsection{Circular orbits and the Lyapunov exponent}
\label{subsec:circular_orbits_LE}

Here we focus on the local radial instability of equatorial circular orbits in the rotating charged Bumblebee background. Combining the coordinate-time radial velocity $\dot r=b_1/a_1$ with the mass-shell condition, we obtain

\begin{equation}
\dot r^{\,2}
=\mathcal G(r;E,J,S)\mathcal H(r;E,J,S),
\label{eq:radial_factorization}
\end{equation}

\noindent with

\begin{equation}
\mathcal H=\mathcal P-\Delta_r, \qquad \mathcal G=\frac{B(r)^2} {(1+\ell)r^2a_1^2},
\label{eq:radial_functions}
\end{equation}

\noindent where $B(r)=1-MS^2/[(1+\ell)r^3]+\bar Q^2S^2/[(1+\ell)r^4]$. The functions $\mathcal P$ and $a_1$ are defined in Eqs.~\eqref{eq:p} and \eqref{eq:a1}, respectively. For fixed background parameters, angular momentum $J$, and signed spin $S$, the conserved energy $E$ and circular-orbit radius $r_0$ are obtained by solving

\begin{equation}
\mathcal H(r_0)=0, \qquad \mathcal H'(r_0)=0.
\label{eq:regular_circular_conditions}
\end{equation}

\noindent These conditions identify candidate circular orbits on regular branches for which $\mathcal G$ is smooth and $0<\mathcal G(r_0)<\infty$. We restrict attention to orbits outside the event horizon, $r_0>r_+$, with $D_S(r_0)\neq0$ and $a_1(r_0)\neq0$. Physical admissibility further requires a future-directed timelike worldline and the validity of the test-particle and pole--dipole approximations, with $m/M\ll1$ and the characteristic spin length small compared with the local curvature scale.

Following the coordinate-time effective-potential approach \cite{JLLL,CMBWZ,CHL}, we write the
radial equation as $\frac{1}{2}m\dot r^{\,2}+\mathcal V_{\rm eff}=0$, where the effective potential is $\mathcal V_{\rm eff}=-m\mathcal G\mathcal H/2$. For a small radial perturbation
$r(t)=r_0+\epsilon(t)$, with $|\epsilon|\ll r_0$, the linearized radial equation takes the form

\begin{equation}
\ddot\epsilon-\lambda^2\epsilon\simeq0,
\label{eq:linear_radial_perturbation}
\end{equation}

\noindent where the squared LE is

\begin{equation}
\begin{aligned}
\lambda^2 =-\frac{\mathcal V_{\rm eff}''(r_0)}{m} =\frac{B(r_0)^2}
{2(1+\ell)r_0^2a_1(r_0)^2} \left(\mathcal P''(r_0)-2\right).
\end{aligned}
\label{eq:LE_simplified}
\end{equation}

\noindent The last equality follows from $\mathcal H''=\mathcal P''-\Delta_r''$ and $\Delta_r''=2$. All radial derivatives are evaluated at fixed $E$, $J$, $S$, and background parameters, retaining the radial dependence of $p_t$ and $p_\phi$. For regular circular orbits, $\lambda^2>0$ indicates local radial instability, with $\epsilon(t)\propto e^{\lambda t}$.

\subsection{Effective KN correspondence: orbital dynamics and limitations}
\label{subsec:effective_KN_correspondence}

We now examine whether the momentum correspondence established in
Eqs.~\eqref{eq:covariant_momentum_mapping}--\eqref{eq:radial_momentum_mapping_full}
extends to the reduced coordinate-time dynamics of neutral spinning particles. Throughout this subsection, we consider matched effective background parameters $(M,\bar a,\bar Q)$ and conserved quantities $(E,J)$, with the signed spins related by Eq.~\eqref{eq:spin_mapping}. The analysis is restricted to regular equatorial configurations within the MPD--TD pole--dipole approximation, with $1+\ell>0$.

\paragraph{Dynamical correspondence.}

For matched effective background parameters, conserved quantities, and rescaled signed spins,
the equality of the temporal--azimuthal metric components and covariant momenta implies
$\mathcal P_{\rm BB}(r)=\mathcal P_{\rm KN}(r)$, as follows from Eq.~\eqref{eq:p}. Since $\Delta_r$ is identical in both backgrounds, Eq.~\eqref{eq:radial_functions} gives

\begin{equation}
\mathcal H_{\rm BB}(r) =\mathcal H_{\rm KN}(r).
\label{eq:effective_H_identity}
\end{equation}

The spin mapping \eqref{eq:spin_mapping} implies $S_{\rm BB}^2=(1+\ell)S_{\rm KN}^2$, and hence $B_{\rm BB}(r)=B_{\rm KN}(r)$. Together with the momentum correspondence and the identical effective background parameters entering $\mathcal A_1$, $\mathcal B_1$ and
$\mathcal C_1$, Eqs.~\eqref{eq:a1}--\eqref{eq:b2} yield $a_{1,\rm BB}=a_{1,\rm KN}$, $a_{2,\rm BB}=a_{2,\rm KN}$, $b_{1,\rm BB}=\frac{b_{1,\rm KN}}{\sqrt{1+\ell}}$ and $b_{2,\rm BB}=b_{2,\rm KN}$. Consequently, for corresponding regular configurations on matching radial branches,
the coordinate velocities satisfy

\begin{equation}
\begin{aligned}
\left(\frac{dr}{dt}\right)_{\rm BB} &=\frac{1}{\sqrt{1+\ell}} \left(\frac{dr}{dt}\right)_{\rm KN},\\
\Omega_{\rm BB} &=\Omega_{\rm KN},
\end{aligned}
\label{eq:coordinate_velocity_mapping}
\end{equation}

\noindent where $\Omega=d\phi/dt$. Furthermore, Eq.~\eqref{eq:radial_functions} gives

\begin{equation}
\mathcal G_{\rm BB}(r) =\frac{\mathcal G_{\rm KN}(r)}{1+\ell}.
\label{eq:effective_G_identity}
\end{equation}

\noindent Together with Eq.~\eqref{eq:effective_H_identity}, this relation reproduces the radial velocity scaling directly from the factorized radial equation \eqref{eq:radial_factorization}.

The velocity correspondence also preserves the
timelike character of the matched trajectories.
Defining
\begin{equation}
\mathcal T
=g_{tt}+2g_{t\phi}\Omega+g_{\phi\phi}\Omega^2
+g_{rr}\left(\frac{dr}{dt}\right)^2,
\label{eq:timelike_mapping_def}
\end{equation}
the metric and velocity relations imply
\begin{equation}
\mathcal T_{\rm BB}=\mathcal T_{\rm KN}.
\label{eq:timelike_mapping}
\end{equation}
Thus, corresponding regular trajectories are
timelike simultaneously. With future-directed
proper-time parametrization, their four-velocity
components satisfy
\begin{equation}
u^t_{\rm BB}=u^t_{\rm KN},\qquad
u^\phi_{\rm BB}=u^\phi_{\rm KN},\qquad
u^r_{\rm BB}
=\frac{u^r_{\rm KN}}{\sqrt{1+\ell}}.
\label{eq:four_velocity_mapping}
\end{equation}
These relations apply to matched equatorial motions
and do not imply an isometry between the backgrounds.

\paragraph{Circular-orbit instability.}

For matched effective background parameters, conserved quantities, and rescaled signed spins,
the identity $\mathcal H_{\rm BB}(r)=\mathcal H_{\rm KN}(r)$ implies that corresponding regular circular orbits satisfy the same conditions \eqref{eq:regular_circular_conditions}.
Provided that $0<\mathcal G_{\rm BB}(r_0)<\infty$ and $0<\mathcal G_{\rm KN}(r_0)<\infty$,
the two systems share the same circular-orbit radius and radial mass-shell curvature
$\mathcal H''(r_0)$. Using Eqs.~\eqref{eq:LE_simplified} and \eqref{eq:effective_G_identity},
we obtain

\begin{equation}
\lambda_{\rm BB}^{2}=\frac{\lambda_{\rm KN}^{2}}{1+\ell}.
\label{eq:effective_LE_mapping}
\end{equation}

In the nonrotating limit $a=0$, the effective KN background reduces to an RN spacetime with
$\bar Q^2=2(1+\ell)Q_0^2/(2+\ell)$. The radial metric relation $g_{rr}^{\rm BB} = (1+\ell) g_{rr}^{\rm RN}$ remains valid, and the equatorial MPD--TD correspondence is retained under the signed-spin rescaling $S_{\rm RN}=S_{\rm BB}/\sqrt{1+\ell}$. The corresponding squared LEs therefore satisfy
\begin{equation}
\lambda_{\rm BB}^2
=\frac{\lambda_{\rm RN}^2}{1+\ell}.
\end{equation}

\paragraph{Limits of the trajectory correspondence.}

Although the corresponding systems have identical coordinate angular velocities, their radial
coordinate velocities differ. For matched trajectory segments with nonzero radial velocity,
Eq.~\eqref{eq:coordinate_velocity_mapping} implies

\begin{equation}
\left(\frac{d\phi}{dr}\right)_{\rm BB}=\sqrt{1+\ell}\left(\frac{d\phi}{dr}\right)_{\rm KN}.
\label{eq:azimuthal_radial_mapping}
\end{equation}

\noindent Integrating over corresponding segments with the same radial endpoints and orientation gives

\begin{equation}
\Delta\phi_{\rm BB}=\sqrt{1+\ell}\,\Delta\phi_{\rm KN},
\label{eq:accumulated_azimuth_mapping}
\end{equation}

\noindent where $\Delta\phi$ denotes the accumulated coordinate azimuthal angle. Thus, the
corresponding trajectories generally trace different curves in the $(r,\phi)$ plane.

For unstable circular orbits with $\Omega\neq0$ and $\lambda>0$, we define the dimensionless radial perturbation-growth exponent over one azimuthal period $T_\phi=2\pi/|\Omega|$ as

\begin{equation}
\mathcal N=\frac{2\pi\lambda}{|\Omega|}.
\label{eq:orbital_growth_parameter}
\end{equation}

\noindent Using Eqs.~\eqref{eq:coordinate_velocity_mapping} and \eqref{eq:effective_LE_mapping}, we obtain

\begin{equation}
\mathcal N_{\rm BB}=\frac{\mathcal N_{\rm KN}}{\sqrt{1+\ell}}.
\label{eq:orbital_growth_mapping}
\end{equation}

\noindent Hence, despite having identical coordinate angular velocities and azimuthal periods,
corresponding unstable circular orbits exhibit different radial perturbation-growth factors, $e^{\mathcal N_{\rm BB}}$ and $e^{\mathcal N_{\rm KN}}$, within the linear approximation.

The correspondence holds for matched effective background parameters, conserved quantities, and rescaled signed spins within the MPD--TD pole--dipole approximation. It does not establish a global isometry or imply the equivalence of all observables. In particular, varying $\ell$ at fixed original parameters $(a,Q_0,S_{\rm BB})$ changes the effective background parameters and the corresponding normalized spin, so the scaling relations derived at fixed effective parameters cannot be interpreted as simple scaling laws for such parameter variations. Moreover, extending the correspondence to charged particles requires a separate analysis of their coupling to the background electromagnetic potential. These results remain subject to the validity of the adopted Bumblebee background and the pole--dipole approximation.

\paragraph{Lorentz-violating response at fixed original parameters.}

The preceding correspondence is established for matched effective background parameters,
conserved quantities, and rescaled signed spins. We now consider varying the Lorentz-violating
parameter $\ell$ while keeping the original parameters $(M,a,Q_0,J,S_{\rm BB})$ fixed. The corresponding effective parameters vary as

\begin{equation}
\bar a(\ell)=a\sqrt{1+\ell}, \qquad \bar Q^2(\ell) =\frac{2(1+\ell)}{2+\ell}Q_0^2,
\qquad S_{\rm KN}(\ell)=\frac{S_{\rm BB}}{\sqrt{1+\ell}}.
\label{eq:fixed_original_parameter_mapping}
\end{equation}

\noindent Here, $S_{\rm KN}(\ell)$ denotes the signed spin assigned to the corresponding effective KN system.

For each value of $\ell$, the circular-orbit conditions determine the energy $E(\ell)$ and radius $r_0(\ell)$ along the selected orbital branch. For regular unstable circular orbits with $\Omega\neq0$, the dimensionless radial growth exponent satisfies

\begin{equation}
\mathcal N_{\rm BB}(\ell)=\frac{2\pi}{\sqrt{1+\ell}}\frac{\lambda_{\rm KN}(\ell)}
{|\Omega_{\rm KN}(\ell)|},
\label{eq:fixed_original_growth}
\end{equation}

\noindent where the effective KN quantities are evaluated at the corresponding circular orbit for each $\ell$. Taking the logarithmic derivative along a smooth orbital branch gives

\begin{equation}
\frac{d\ln\mathcal N_{\rm BB}}{d\ell}=-\frac{1}{2(1+\ell)} +\frac{d\ln\lambda_{\rm KN}}{d\ell}
-\frac{d\ln|\Omega_{\rm KN}|}{d\ell}.
\label{eq:fixed_original_growth_response}
\end{equation}

\noindent The total derivatives include the implicit dependence of the circular-orbit energy and radius on $\ell$. The first term represents the explicit radial rescaling, whereas the remaining terms describe the response of the effective KN orbital instability and angular velocity to the parameter variation. Thus, at fixed original parameters, the Lorentz-violating response of the radial growth exponent is governed by both the explicit rescaling and the induced changes in the effective orbital dynamics. The relation holds along smooth branches of regular unstable circular orbits within the domain of validity of the adopted background and MPD--TD approximation.

The corresponding LE scaling, together with the surface-gravity relation derived from the background geometry, also provides a basis for comparing the local orbital-instability and horizon surface-gravity scales. This comparison is discussed in Sec.~\ref{sec5.2}.

\section{Physical interpretation and applications}
\label{sec5}

We now examine the geometric origin and physical implications of the effective KN correspondence, with particular emphasis on the distinction between analytical correspondence and dynamical equivalence. We then apply the correspondence to the comparison between local orbital instability and black hole surface gravity.

\subsection{Geometric origin and dynamical implications}
\label{sec5.1}

The effective KN correspondence originates from the specific radial structure of the Bumblebee metric. Analytical constructions of Bumblebee geometries have shown that Lorentz-violating contributions can, under specific assumptions, be represented by additive modifications of general-relativistic background metrics~\cite{Poulis2022,Ovcharenko2026}. For matched effective parameters $(M,\bar a,\bar Q)$, the metric components in the $t$--$\phi$ sector
coincide, whereas the radial components differ by the constant factor $1+\ell$. This structure permits a common algebraic description of the temporal and azimuthal dynamics, while radial evolution acquires an additional scaling factor. The resulting correspondence is an algebraic
property of the adopted slowly rotating charged Bumblebee metric and does not imply a global
isometry or an equivalence of the underlying field configurations.

For spinless particles, this structure leads directly to a common radial mass-shell function,
with the radial kinetic term modified by the Lorentz-violating factor. For spinning particles,
the situation is more subtle because spin--curvature coupling involves derivatives of the metric and the four-momentum is generally not parallel to the four-velocity. The correspondence survives within the equatorial MPD--TD dynamics only after the spin-dependent terms are consistently matched through the signed-spin rescaling. Its existence is therefore a nontrivial property of the reduced dynamical equations, rather than a consequence of metric similarity alone.

The physical significance of the correspondence lies in the distinction between the preservation of selected circular-orbit properties and the inequivalence of radial evolution. The Lorentz-violating deformation leaves the matched coordinate angular velocities unchanged but modifies radial motion, accumulated azimuthal angles, and perturbation growth over an orbital period. Thus, the effective KN correspondence provides a common analytical description of these quantities without implying complete equivalence of the underlying dynamics.

\subsection{Application to the chaos-bound criterion}
\label{sec5.2}

The effective KN correspondence also determines how the local radial instability of circular test-particle orbits compares with the black hole surface gravity. Motivated by earlier studies
of classical orbital instability and chaos bounds in black hole spacetimes \cite{HT,ZLL}, we introduce the classical chaos-bound indicator \cite{LG1,KG2022}

\begin{equation}
\Delta_{\lambda\kappa}
=\lambda^2-\kappa^2,
\label{eq:chaos_indicator}
\end{equation}

\noindent \noindent where $\lambda$ is the coordinate-time radial LE of an unstable circular orbit, and $\kappa$ is the black hole surface gravity. For matched Bumblebee and effective KN configurations, the LE and surface gravity satisfy

\begin{equation}
\lambda_{\rm BB}^2
=\frac{\lambda_{\rm KN}^2}{1+\ell},
\qquad
\kappa_{\rm BB}^2
=\frac{\kappa_{\rm KN}^2}{1+\ell},
\label{eq:chaos_scale_correspondence}
\end{equation}

\noindent where the surface gravities are defined using the same normalization of the
stationary Killing time. Consequently,

\begin{equation}
\Delta_{\lambda\kappa}^{\rm BB}
=\frac{\Delta_{\lambda\kappa}^{\rm KN}}
{1+\ell}.
\label{eq:chaos_bound_application_mapping}
\end{equation}

\noindent Since $1+\ell>0$, the sign of the indicator is preserved between corresponding configurations. Thus, matched unstable circular orbits lie on the same side of the threshold \(\lambda=\kappa\). For nonextremal black holes with $\kappa\neq0$, the dimensionless
ratio is also invariant:

\begin{equation}
\frac{\lambda_{\rm BB}^2}{\kappa_{\rm BB}^2}
=\frac{\lambda_{\rm KN}^2}{\kappa_{\rm KN}^2}.
\label{eq:chaos_ratio_application}
\end{equation}

\noindent The spinless correspondence is recovered by setting $S_{\rm BB}=S_{\rm KN}=0$. This sign preservation applies to matched configurations and should be distinguished from variations of $\ell$ at fixed original parameters $(M,a,Q_0,J,S_{\rm BB})$. In the latter case, the effective rotation, charge, and rescaled signed spin vary with $\ell$, potentially changing the corresponding circular orbit and instability scale. Where the correspondence remains valid, a sign change in $\Delta_{\lambda\kappa}^{\rm BB}$ along a physically admissible family must therefore be accompanied by a sign change in the indicator of the corresponding effective KN configuration, rather than arising from the positive rescaling factor alone.

These results establish the preservation of the sign of the classical chaos-bound indicator
under the effective KN correspondence, despite the difference in the radial geometries. The conclusion applies to matched equatorial configurations within the validity of the adopted background and the MPD--TD pole--dipole approximation \cite{KNS2026}. The LE considered here characterizes local classical instability, and its comparison with the surface gravity does not constitute a direct test of the quantum many-body chaos bound \cite{MSS}.

\section{Conclusions}
\label{sec6}

We have investigated the equatorial dynamics of neutral spinless and spinning test particles in a rotating charged BKN black hole background and established an analytical correspondence with their dynamics in an effective KN background. The two metrics share the same temporal--azimuthal sector when their effective parameters are matched, whereas their radial components differ by a factor of $1+\ell$. This geometric relation also extends to the equatorial MPD--TD dynamics through a rescaling of the signed particle spin.

The correspondence preserves the circular-orbit conditions, orbital radii, and coordinate angular velocities for matched conserved quantities and rescaled spins. The radial coordinate velocities and squared LEs, however, acquire factors depending on the Lorentz-violating parameter. As a result, corresponding noncircular trajectories generally accumulate different azimuthal angles between the same radial endpoints, and radial perturbations of unstable circular orbits grow by different factors over one azimuthal period. Despite these differences, the squared LE and squared surface gravity obey the same positive rescaling, leaving the sign of the chaos-bound indicator unchanged. This analytical correspondence does not imply a global equivalence of the two spacetimes. In particular, varying the Lorentz-violating parameter while keeping the original background and particle parameters fixed also changes the effective KN configuration, and the orbital response cannot be reduced to a radial rescaling alone. Our results apply within the adopted rotating Bumblebee background and the equatorial MPD--TD pole--dipole approximation. The instability criterion discussed here concerns local classical orbital dynamics and should not be interpreted as a direct test of quantum many-body chaos.

\end{document}